\newcommand{\gpm}[3]{$#1^{+#2}_{-#3}$}
\def\l{$\lambda$}
\def\degr{\hbox{$^\circ$}}

\def\arcsec{\hbox{$^{\prime\prime}$}}

\def\farcs{\hbox{$.\!\!^{\prime\prime}$}}  
\def\lsim{\mathrel{\hbox{\rlap{\lower.55ex \hbox {$\sim$}}\kern-.0em
\raise.4ex \hbox{$<$}}}} 
\def\gsim{\mathrel{\hbox{\rlap{\lower.55ex \hbox {$\sim$}}\kern-.0em
\raise.4ex \hbox{$>$}}}} 
\def\lya{Ly$\alpha$}

\def\9{GRB\,991216}
\def\1{GRB\,011211}
\def\2{GRB\,021211}
\def\subsun{\mbox{$_{\odot}$}}
\def\ion#1#2{#1$\;${\small\rm\@Roman{#2}}\relax}
\def\l{$\lambda$}
\def\kms{km s$^{-1}$}

\documentclass{aa}
\usepackage{natbib}
\usepackage{aas_macros}
\usepackage{graphicx}

\bibpunct{(}{)}{;}{a}{}{,}

\begin{document}
\title{Low-resolution VLT spectroscopy of GRBs 991216, 011211 and
021211\thanks{Based on observations collected at the European Southern
Observatory, Chile; proposals no. 64.H-0313, 165.H-0464, 70.D-0523}}

%\subtitle{}

\author{P.~M. Vreeswijk\inst{1,2}
  \and
  A. Smette\inst{1,3,4}
  \and
  A. S. Fruchter\inst{5}
  \and
  E. Palazzi\inst{6}
  \and
  E. Rol\inst{2,7}
  \and
  R.~A.~M.~J. Wijers\inst{2,8} 
  \and
  C. Kouveliotou\inst{9}
  \and
  L. Kaper\inst{2}
  \and
  E. Pian\inst{6,10}
  \and
  N. Masetti\inst{6}
  \and
  F. Frontera\inst{11,6}
  \and
  J. Hjorth\inst{12}
  \and
  J. Gorosabel\inst{13}
  \and
  L. Piro\inst{14}
  \and
  J.~P.~U. Fynbo\inst{12}
  \and
  P. Jakobsson\inst{12}
  \and
  D. Watson\inst{12}
  \and
  P.~T. O'Brien\inst{7}
  \and
  C. Ledoux\inst{1}
}

\offprints{pvreeswi@eso.org}
   
\institute{European Southern Observatory, Alonso de C\'ordova 3107,
  Casilla 19001, Santiago 19, Chile
  \and
  Astronomical Institute `Anton Pannekoek', University of Amsterdam \&
  Center for High Energy Astrophysics, Kruislaan 403, 1098 SJ
  Amsterdam, The Netherlands
  \and
  Institut d'Astrophysique et de G\'eophysique, Universit\'e de
  Li\`ege, All\'ee du 6 Ao\^ut, 17, B-4000 Li\`ege, Belgium
  \and
  Chercheur qualifi\'e, F.N.R.S., Belgium
  \and
  Space Telescope Science Institute, 3700 San Martin Drive, Baltimore,
  MD 21218, USA
  \and
  INAF - Istituto di Astrofisica Spaziale e Fisica Cosmica, Sezione di
  Bologna, Via Gobetti 101, I-40129 Bologna, Italy
  \and
  Department of Physics and Astronomy, University of Leicester,
  University Road, Leicester, LE1 7RH, UK
  \and
  Department of Physics and Astronomy, SUNY Stony Brook, NY
  11794-3800, USA 
  \and
  National Space Science Technology Center, NASA/MSFC, XD-12, 320
  Sparkman Drive, Huntsville, AL 35805, USA
  \and
  INAF - Osservatorio Astronomico di Trieste, Via G.B. Tiepolo 11,
  I-34131 Trieste, Italy
  \and
  Dipartimento di Fisica, Universit\`a di Ferrara, via Paradiso 12,
  I-44100 Ferrara, Italy
  \and
  Dark Cosmology Centre, Niels Bohr Institute, University of
  Copenhagen, Juliane Maries Vej 30, DK-2100 Copenhagen {\O}, Denmark
  \and
  Instituto de Astrof\'{\i}sica de Andaluc\'{\i}a, CSIC, Apartado
  3004, 18080 Granada, Spain.
  \and
  INAF - Istituto di Astrofisica Spaziale e Fisica Cosmica, Sezione di
  Roma, via Fosso del Cavaliere 100, I-00133 Roma, Italy}

\date{\today}
   
\authorrunning{Vreeswijk et al.}
\titlerunning{VLT spectroscopy of GRBs 991216, 011211 and 021211}
   
\abstract{We present low-resolution VLT spectroscopy of the afterglow
  of the gamma-ray bursts (GRBs) 991216, 011211 and 021211. Our
  spectrum of \9\ is the only optical spectrum for this afterglow. It
  shows two probable absorption systems at $z=0.80$ and $z=1.02$,
  where the highest redshift most likely reflects the distance to the
  host galaxy. A third system may be detected at $z=0.77$. HST imaging
  of the field, obtained 4 months after the burst, has resulted in the
  detection of two amorphous regions of emission, one at the projected
  afterglow position, and the other 0\farcs6 away. The spectrum shows
  a depression in flux in between 4000~\AA\ and 5500~\AA. This could
  be the result of a 2175~\AA-type extinction feature in the host of
  \9, but at a rather red wavelength of 2360~\AA. If this
  interpretation is correct, it is the first time the extinction
  feature is seen in a GRB afterglow spectrum. It is centered at a
  wavelength similar to that of the ultra-violet (UV) bumps inferred
  from observations of a few UV-strong, hydrogen-poor stars in the
  Galaxy.  All significant absorption lines (except for one) detected
  in the spectrum of \1\ are identified with lines originating in a
  single absorption system at $z=2.142\pm0.002$, the redshift of the
  \1\ host galaxy. We also detect the \lya\ absorption line in the
  host, to which we fit a neutral hydrogen column density of
  log~$N$(\ion{H}{i})=$20.4\pm0.2$, which indicates that it is a
  damped \lya\ system. Using a curve-of-growth analysis, we estimate
  the Si, Fe and Al metallicity at the \1\ redshift to be
  [Si/H]=\gpm{-0.9}{0.6}{0.4}, [Fe/H]=$-1.3\pm0.3$, and
  [Al/H]=\gpm{-1.0}{0.5}{0.3}. For \2, we detect a single emission
  line in a spectrum obtained tens of days after the burst, which we
  identify as [\ion{O}{ii}]~\l 3727 at $z=1.006$. The corresponding
  unobscured [\ion{O}{ii}] star-formation rate is
  1.4~M\subsun\ yr$^{-1}$. \keywords{Gamma rays: bursts -- Galaxies:
    abundances, distances and redshifts}} \maketitle
%
%________________________________________________________________

\section{Introduction}
\label{sec:introduction}

Spectroscopy of the afterglows of long-duration Gamma-Ray Bursts
(GRBs) has been essential for our understanding of the physical
mechanism that produces these powerful explosions. Following the
discovery of GRB afterglows
\citep{1997Natur.387..783C,1997Natur.386..686V}, the first redshift
determination by \citet{1997Natur.387..878M} provided conclusive
evidence that the origin of the long-duration class of GRBs
\citep{1993ApJ...413L.101K} is cosmological. Obviously, redshifts are
required to deduce most meaningful quantities, such as the GRB
energetics
\citep{frailbeaming,2001AJ....121.2879B,2004ApJ...616..331G}, the
brightness of any underlying supernova component
\citep[e.g.][]{1999Natur.401..453B}, and properties of the GRB host
galaxies \citep[e.g.][]{1999ApJ...520...54H}. So far, optical
redshifts have been secured for four dozen GRBs, either through
absorption-line spectroscopy of their bright early afterglow
\citep[e.g.][]{1997Natur.387..878M}, or through the detection of
host-galaxy emission lines \citep[e.g.][]{1999Natur.398..389K}.  The
average observed GRB redshift is $z=1.3$
\citep[see][]{030429dla,bergerswift}, with the highest being
GRB~050904 at $z=6.29$ \citep{2005GCN..3937....1K}.

Apart from the vital redshift, spectroscopic afterglow observations
have also shown that GRB hosts are actively star-forming galaxies
\citep[e.g.][]{1998ApJ...508L..17D,2001ApJ...546..672V}, that
the afterglow is often situated behind a very large neutral hydrogen
column \citep[e.g.][]{hjorth020124,vrees030323} with relatively high
metal column densities and dust depletions compared to damped
\lya\ systems observed along QSO sight lines 
\citep{2003ApJ...585..638S,2004ApJ...614..293S}, and
have provided evidence for high-velocity outflows \citep[up to 3000
\kms,][]{2002A&A...396L..21M,2003ApJ...588..387S,2003ApJ...595..935M},
presumably caused by the wind of the GRB massive-star
progenitor. Finally, spectroscopic monitoring of GRB\,030329/SN2003dh
\citep{stanek,hjorth030329} has revealed a remarkable similarity with the
spectral evolution of GRB\,980425/SN1998bw
\citep{1998Natur.395..670G,2001ApJ...555..900P}, confirming the
connection between GRBs and supernova explosions. This strongly
suggests that long-duration GRBs are produced by collapsing massive
stars \citep[see][]{1993ApJ...405..273W,2001ApJ...550..410M}.

The GRB Afterglow Collaboration at ESO (GRACE) has an on-going program
at the European Southern Observatory (ESO) to perform spectroscopic
observations of GRB afterglows. In this paper, we present
low-resolution Very Large Telescope (VLT) spectroscopy of three GRB
afterglows: 991216, 011211 and 021211. For
\9\ the spectroscopy is complemented by {\it Hubble Space Telescope} (HST)
imaging.  The organisation of this paper is as follows. After a
description of the observations, data reduction and spectral analysis
in Sect.~\ref{sec:observations}, each GRB afterglow is presented in a
separate section: \9\ in Sect.~\ref{sec:grb991216}, \1\ in
Sect.~\ref{sec:grb011211} and \2\ in Sect.~\ref{sec:grb021211}.  Each
of these sections starts with an introduction on the afterglow,
followed by our results. We briefly conclude in
Sect.~\ref{sec:conclusions}.

\begin{table*}[tbp]
  \centering
  \caption[]{Log of VLT spectroscopic observations.}\label{tab:obs}
  \null\vspace{-1.0cm}
  $$
  \begin{array}{ccrcllccclcc}
    \hline
    \hline
    \noalign{\smallskip}
    \rm GRB &
    \rm UT \, date ^{\mathrm{a}} &
    \rm \Delta T &
    \rm instr. &
    \rm grism \, (filter) &
    \rm coverage ^{\mathrm{b}} &
    \rm slit width &
    \rm resolution ^{\mathrm{c}}  &
    \rm dispersion &
    \rm exptime &
    \rm seeing &
    \rm airmass \\
    &
    &
    \rm (days) &
    &
    &
    \rm (\AA) &
    (\arcsec) &
    \rm (\AA) &
    \rm (\AA/pixel) &
    \rm (min) &
    (\arcsec) &
    \\
    \hline
    991216 & \rm 1999\,Dec\,18.159 & 1.49  & \rm FORS1 & \rm 150I (OG590) & 3800$--$9500 & 1 & 24 & 5.3 & 6\times10^{\mathrm{d}}& 0.6 & 1.24$--$1.27\\
    011211 & \rm 2001\,Dec\,13.253 & 1.45  & \rm FORS2 & \rm 300V         & 3600$--$8900 & 1 & 10 & 2.6 & 9\times10             & 1.1 & 1.20$--$1.94\\
    021211 & \rm 2002\,Dec\,30.294 & 18.82 & \rm FORS2 & \rm 300V         & 3600$--$9500 & 1 & 11 & 3.2 & 3\times10             & 0.7 & 1.20$--$1.24\\
    \hline 
  \end{array}
  $$
  \begin{list}{}{} 
  \item[$^{\mathrm{a}}$] Start of the first exposure.
  \item[$^{\mathrm{b}}$] Without order sorting filter, the second
  order will start to contaminate the spectrum above $\sim7000$~\AA.
  \item[$^{\mathrm{c}}$] The resolution is approximately constant
  across the entire spectrum.
  \item[$^{\mathrm{d}}$] The first three exposures are taken without order
  sorting filter, followed by three with filter.
  \end{list}
\end{table*}

\section{Observations, data reduction and spectral analysis}
\label{sec:observations}

Table~\ref{tab:obs} shows the log of the VLT spectroscopic
observations.  All spectra have been reduced in the same manner within
IRAF, using tasks within the {\it kpnoslit} package. After overscan
subtraction and flat-fielding, the cosmic rays were removed from the
images using the L.A. Cosmic routine written by
\citet{2001PASP..113.1420V}. The spectra were then optimally extracted
for each 2-dimensional (2-D) image separately, with an extraction
aperture width of 2-4\arcsec.  The wavelength calibration was applied,
again to each spectrum separately, using an HeNeAr lamp spectrum that
was taken in the morning after the science observations. The formal
error in the wavelength calibration fit (order 5) was roughly
0.3~\AA\ for the 150I grism, and 0.2~\AA\ for the 300V grism. The
individual wavelength-calibrated spectra were averaged, and the
corresponding Poisson error spectra, calculated by the {\it apall}
task, were quadratically averaged.

The flux calibration was performed using observations of the standard
HD~49798 for both \9 and \1, and LTT~3218 for \2\footnote{see
  http://www.eso.org/observing/standards/spectra/}.  These standards
were taken with a 5\arcsec\ slit width during the same night as the
GRB science observations. During the night of observations of \9,
HD~49798 was also observed with a slit width of 1\arcsec, i.e. the
same width as that of the GRB afterglow spectra. The nights that
\9\ and \2\ were observed were probably photometric, but the night
that \1\ was observed was definitely not.

For the spectra of \1, we performed a correction for the slit loss,
i.e. the fraction of the surface underneath the spectral profile that
falls outside the slit width. The correction factor was estimated by
fitting a Gaussian profile along the spatial direction (i.e. along the
CCD columns) of the 2-D spectra, averaged over 4 pixels in the
dispersion direction (i.e. averaging 4 columns before performing the
fit). The resulting Gaussian full width at half maximum (FWHM) was
then compared to the slit width to obtain the slit loss along the
dispersion axis. The slit loss profile was then fit with a polynomial
to correct the spectra.  Note that this does not correct for any
colour-dependent slit losses, but both FORS1 and 2 have a linear
atmospheric dispersion compensator (LADC) in the light path, which
minimises any colour-differential slit losses up to a zenith distance
of 45\degr. However, several spectra of
\1\ were taken at an airmass above 1.4 (see Table~\ref{tab:obs}), and
therefore these are likely to suffer from such colour-differential
slit losses. For the \9\ spectra a slit-loss correction was not
necessary, as we flux calibrated with the standard that was taken with
the same slit width as the afterglow spectra. This should in principle
result in a correct flux calibration, provided that the seeing was the
same during the afterglow and standard star observations. This was
more or less the case for \9, with a seeing of 0\farcs6 during the
afterglow observations, and 0\farcs7 during the standard star
observations. For the late-time spectra of \2, where the afterglow
continuum is absent, we did not attempt a slit-loss correction.

Finally, each spectrum was corrected for Galactic extinction, assuming
the $E_{B-V}$ from \citet{1998ApJ...500..525S}. The $E_{B-V}$ values
used are 0.626 for \9, 0.043 for \1, and 0.028 for \2. The flux
calibration, slit-loss correction, and dereddening were also applied
to the combined wavelength-calibrated error spectrum.

The resulting spectra were analysed within IRAF as well. The
equivalent width (EW) of the significant absorption and emission lines
were measured using the {\it splot} routine in IRAF. By eye, we fit
the continuum with a high-order polynomial function. After
normalisation, we measure the EWs with the {\it e} option.  The
Poisson error in the EW is calculated from the formal error spectrum
calculated within IRAF, using the same pixels that were used to
measure the EW. We do not include a contribution in the error from the
uncertainty in the location of the continuum.

\section{GRB~991216}
\label{sec:grb991216}

Following the {\it Burst And Transient Source Experiment} (BATSE)
$\gamma$-ray detection of \9\ \citep[see][]{1999GCN...463....1K}, the
X-ray afterglow was discovered by the {\it Rossi X-ray Timing
Explorer} (RXTE) \citep{Takeshima99}, and the optical/infrared
afterglow by \citet{1999GCN...472....1U}.  The afterglow of \9\ is one
of a number of GRB afterglows \citep[as e.g.  GRB\,990510;
see][]{1999ApJ...523L.121H,2001A&A...372..456P} to show evidence for a
``beaming'' break \citep{1997ApJ...487L...1R}, suggesting that the
gamma-ray, X-ray, and early optical emission was confined into jets
\citep{2000ApJ...543..697H}.  Although the optical through X-ray data
can be explained with a jet fireball model, inclusion of the radio
data calls for more exotic models \citep{2000ApJ...538L.129F}. {\it
Chandra X-ray Observatory} spectral observations of the X-ray
afterglow show two probable emission features \citep[a 4.7$\sigma$
detection at $E\sim3.5$~keV, and a marginal one at
$E\sim4.4$~keV;][]{2000Sci...290..955P}, that are identified with an
iron line and the recombination continuum at a redshift of
$z=1.00\pm0.02$.  These features suggest the presence of 0.01--1
M$_{\odot}$ of iron in the vicinity of the burst
\citep{2000Sci...290..955P,2001ApJ...550L..43V}. However, 
in an independent analysis of the same {\it Chandra} data,
\citet{sako} only find a $\sim2.3\sigma$ fluctuation at a slightly
higher energy (3.8~keV), and do not find any feature near 4.4~keV.

\subsection{Absorption lines and redshift}

\begin{figure*}[t]
  \centering \includegraphics[width=15cm]{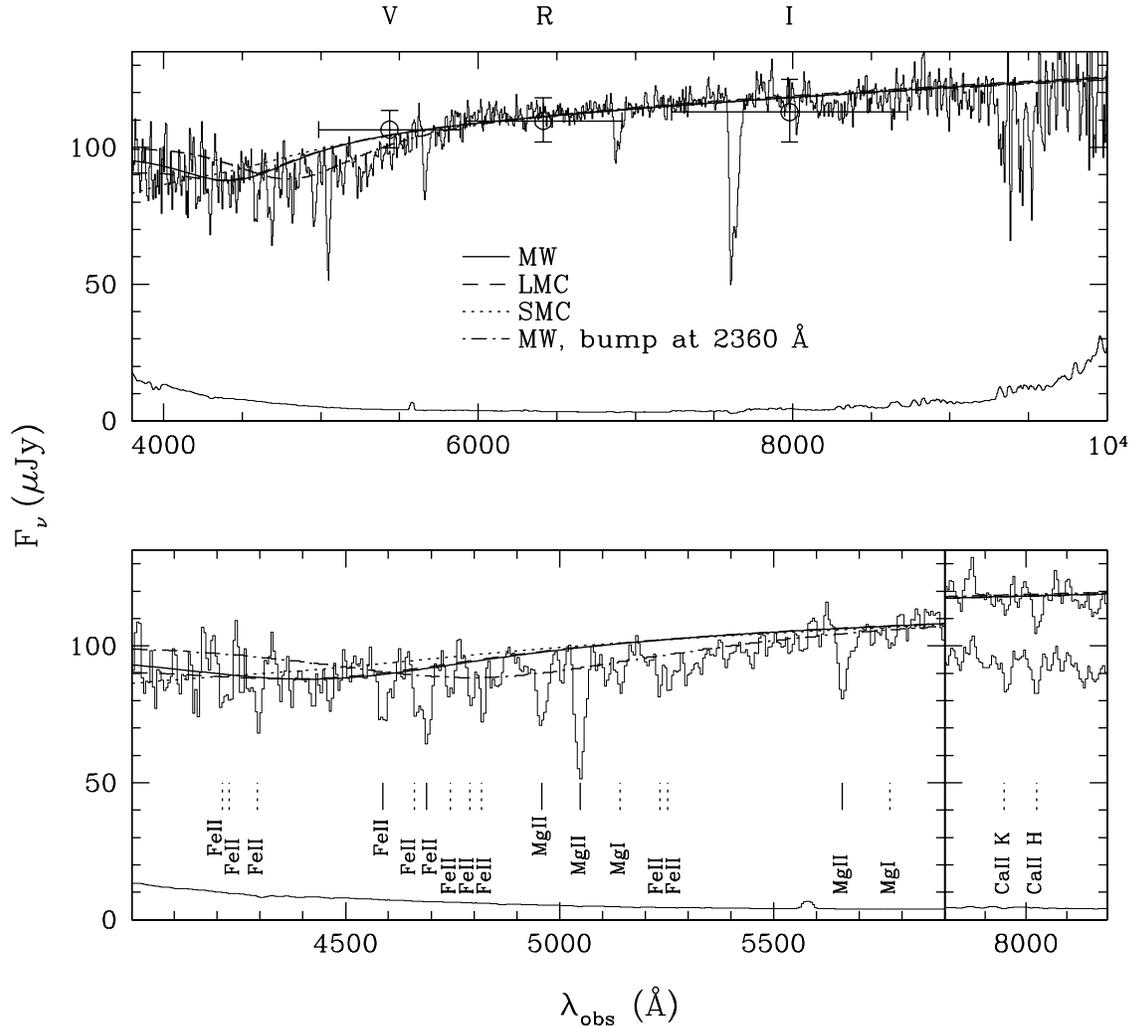}
  \caption{{\bf Top panel:} the optical spectrum of the afterglow of
  \9, taken 1.5 days after the burst. The resolution is approximately
  24~\AA\ across the entire spectrum, which corresponds to a resolving
  power of 250 at 6000~\AA.  The spectrum has not been smoothed. The
  blue and red spectra (respectively without and with the OG590 order
  sorting filter) are connected at 6200~\AA. The two strong absorption
  lines longward of 6000~\AA\ are telluric. The spectrum flux has been
  increased by 10\% to match the $VRI$ photometric points of
  \citet{2000ApJ...543..697H}, indicated with the open circles; the
  horizontal error bars reflect the FWHM of the filter transmission
  profiles. We have fit MW-, LMC- and SMC-type extinction curves to
  the spectrum, shown by the solid (MW), dashed (LMC) and dotted (SMC)
  lines in both panels (see text). The dash-dotted line shows the MW
  fit assuming that the bump is centered at 2360~\AA. {\bf Bottom
  panel}: a blow-up of the top spectrum, over the wavelength ranges
  4000--5900~\AA\ and 7800--8200~\AA. For the latter range, we also
  plot the combined blue and red spectra (offset by 30~$\mu$Jy for
  clarity) below the red-only spectrum. We have identified several
  strong rest-frame UV absorption features, which are indicated with
  short solid lines.  Based on the inferred redshifts of these
  features, there seem to be numerous weaker lines, indicated with
  short dashed lines.  The lower line identification labels correspond
  to the $z=1.02$ system, the middle ones to $z=0.80$ and the top ones
  to $z=0.77$ (see also Table~\ref{tab:lines991216}). For both panels,
  the 1$\sigma$ Poisson error spectrum is also plotted (solid line at
  the bottom).  \label{fig:spectrum991216}}
\end{figure*}

\begin{table}[b]
  \caption[]{Lines detected above 5$\sigma$ in the spectrum of \9.}
  \label{tab:lines991216}
  $$
  \begin{array}{crrcc}
    \hline
    \hline
    \noalign{\smallskip}            
    \lambda_{\rm obs} &
    W_{\rm obs} (\rm\AA) &
    \rm ID &
    z_{\rm abs} \\
    \hline
    4587 &  6.6 \pm 1.3 & \rm Fe${\sc ii}$\,$\l$ 2600 & 0.764 \\
    4689 &  6.2 \pm 1.0 & \rm Fe${\sc ii}$\,$\l$ 2600 & 0.803 \\
    4957 &  7.0 \pm 0.8 & \rm Mg${\sc ii}$\,$\l$ 2800 & 0.770 \\
    5047 & 11.0 \pm 0.7 & \rm Mg${\sc ii}$\,$\l$ 2800 & 0.803 \\
    5662 &  6.6 \pm 0.6 & \rm Mg${\sc ii}$\,$\l$ 2800 & 1.022 \\
    \hline
  \end{array}
  $$
%  \begin{list}{}{}
%  \item[$^{\mathrm{a}}$] 
%  \end{list}
\end{table}

The spectrum of the afterglow of \9\ is shown in
Fig.~\ref{fig:spectrum991216}. The two strong features in the red part
around 6900~\AA\ and 7600~\AA\ are caused by the earth's atmosphere
and are not intrinsic to the GRB host. The apparent lines above
9000~\AA\ are due to bad sky subtraction.  In the blue part of the
spectrum, several significant absorption lines are detected. From
inspection of the standard star spectrum at the same wavelengths we
can conclude that the detected lines are not caused by the instrument,
or by the earth's atmosphere.  The detected lines cannot be identified
with typical interstellar medium (ISM) absorption features of a system
at a single redshift; at least three systems need to be invoked. An
alternative explanation is that the feature around 5700~\AA\ is
Ly$\alpha$ at $z=3.6$, and that the features shortward of it are
Ly$\alpha$ forest lines. We have simulated such a spectrum, and the
lines can appear quite similar to the lines in the observed
spectrum. This is a very unlikely explanation, however, since the
break is much too smooth, and the continuum emission around 4000~\AA\
is only about 10-20\% lower than the level beyond the supposed
break. We therefore discard the latter possibility, and are left with
the most probable explanation: several absorption systems.

Table~\ref{tab:lines991216} lists the identification of the
significant lines in the spectrum of \9. These significant lines are
indicated with the short solid lines in the lower panel of
Fig.~\ref{fig:spectrum991216}. As the spectrum was taken with a very
low-resolution grism (150I), we conservatively use 5$\sigma$ as the
significance threshold.  Typical ISM absorption lines that could be
present at low significance, such as Mg{\sc i}~\l 2852, Ca{\sc ii} K
and H, and other Fe lines, are shown with a short dashed line in the
lower panel of Fig.~\ref{fig:spectrum991216}.

On the basis of the strongest features we infer the possible presence
of three absorption-line systems along the line of sight, with the
following redshifts: $z=0.77$, $z=0.80$ and $z=1.02$. The
identification of the $z=1.02$ system in the blue part of the spectrum
is strengthened by the possible detection of Ca{\sc ii} K (3933~\AA)
and H (3968~\AA) around 8000~\AA, respectively at $z=1.021$ and
$z=1.022$. In order to increase the signal-to-noise ratio (S/N) around
the possible Ca lines, we have added the blue spectrum (taken without
the order sorting filter) to the red spectrum (taken with the OG590
order sorting filter), even though the blue spectrum can be
contaminated by its second order above $\sim7000$~\AA. This second
order appears with twice the resolution and at twice the wavelength of
the first order.  However, if the two features around 8000~\AA\ were
due to this contamination, their first order counterparts would be
located at 3976~\AA\ and 4009~\AA, where no lines are detected.  The
resulting combined spectrum around the Ca lines is also shown in
Fig.~\ref{fig:spectrum991216}, offset downward by 30 $\mu$Jy from the
red-only spectrum for clarity. We measure the following equivalent
widths for the two features in the combined spectrum: $W_{\rm
obs}$(Ca{\sc ii}~K~\l 3933) = ($1.7\pm0.4$)~\AA\ and $W_{\rm
obs}$(Ca{\sc ii}~H~\l 3968) = ($2.4\pm0.4$)~\AA. The expected ratio of
this doublet on the basis of the oscillator strengths, $\frac{\rm
CaII~K~\lambda 3933}{\rm CaII~H~\lambda 3968}$, is roughly two if the
lines are not saturated, but can be unity if they are saturated.  The
observed ratio, $0.7\pm0.2$, indicates that these lines are saturated.

A redshift of $z=1.02$ for \9\ is consistent with the inferred
$z=1.00\pm0.02$ iron line detection in the {\it Chandra} spectrum of
the X-ray afterglow of \9 \citep{2000Sci...290..955P}.  However, we
note that in an independent analysis of the {\it Chandra} data,
\citet{sako} find this iron line to be not significant.  The $z=1.02$
system that we detect in our optical spectrum is most likely caused by
the \9\ host-galaxy ISM.  In all cases for which both GRB absorption
and emission lines have been detected, the (most distant) absorption
system is at the same redshift as the emission lines from the presumed
host galaxy \citep[e.g. for GRB\,980703 and
GRB\,021004;][]{1998ApJ...508L..17D,2003ApJ...595..935M}.  Moreover,
if the detection of Mg{\sc i} at $z=1.02$ is real, this suggests a
dense environment at this redshift, probably the ISM of the host
galaxy. The probable detection of the Ca{\sc ii} doublet supports this
hypothesis, since this line is believed to require column densities
that are typically an order of magnitude larger than seen in Mg{\sc
ii} absorbers \citep{1991MNRAS.251..649B,1992ApJ...399..373C}. We note
that the $z=0.80$ system shows the strongest Mg{\sc ii} and Fe{\sc ii}
features.

\subsection{The shape of the spectral continuum}
\label{sec:991216shape}

The continuum flux level of optical afterglows, measured from
spectroscopy or broad-band photometry, is generally well described by
a single power law, with a slope of the order $\beta=-0.9$ (with
F$_{\nu}\propto \nu^{\beta}$)
\citep[for a high S/N example,
see][]{2004ApJ...614..293S}. This is consistent with the fireball
afterglow theory \citep[e.g.][]{1998ApJ...497L..17S}.  Any departure
from the pure power law is usually ascribed to extinction caused by
the host-galaxy ISM \citep[see][]{1998Natur.393...43R}. The
flux-calibrated spectrum of \9\ contains two features that distinguish
it from ``normal'' afterglow spectra: its shallow slope ($\beta=-0.19$
over the range 6000-9300~\AA), and an apparent depression around
4700~\AA.

In order to verify our absolute flux calibration, we first compare the
spectrum with the photometry of \citet{2000ApJ...543..697H}. Their
$VRI$ measurements are shown in Fig.~\ref{fig:spectrum991216}; the
horizontal error bars represent the approximate FWHM of the filter
transmission curves. In absolute terms, the spectral flux level is
roughly 10\% below that of the photometry; we have scaled the spectrum
in Fig.~\ref{fig:spectrum991216} upward with this amount. Relatively,
however, our flux calibration is in good agreement with the $VRI$
photometry. There are no published $U$ or $B$ photometry data to
compare with. A cause for concern is that the spectrophotometric
standard that was used for the \9\ flux calibration, HD~49798, has not
been calibrated in the optical regime with an observed spectrum, but
rather with a model extension from the observed 1150-3200~\AA\ range
\citep[see][]{bohlin}. The same standard was also observed for \1\
(see Sect.~\ref{sec:011211spectroscopy}).  Comparison of our \1\
spectrum (see Fig.~\ref{fig:spectrum011211}) with the photometry from
\citet{2003A&A...408..941J} indicates that the absolute flux
calibration of the \1\ spectrum is too high by 70\%. Indeed, the night
when this spectrum was taken was definitely not photometric
(alternating thin and thick cloud conditions).  However, after scaling
the spectral flux down by this 70\%, the relative spectral flux
calibration of our \1\ spectrum is in good agreement with the
photometric calibration. In the B band, the spectral flux is
$\sim10$\% higher than the photometry. If this is due to an error in
the model flux calibration of HD~49798, and we were to apply a
correction to the \9\ spectrum based on this difference, it would only
make the \9\ flux depression more pronounced.

Also, during the night that the \9\ afterglow spectra were taken, the
standard HD~49798 was observed with a slit width of both 1\arcsec\ (in
long-slit mode, or LSS) and 5\arcsec\ (in multi-object mode, or MOS;
see the FORS manual\footnote{see http://www.eso.org/instruments/fors/}
for details). This allows us to check whether the depression could be
due to colour-differential, or wavelength-dependent slit losses.
Flux-calibration of the \9\ spectra with the 5\arcsec\ standard
results in an absolute offset of about 13\% with the
1\arcsec-flux-calibrated spectra, with a lower flux level for the
5\arcsec-calibrated spectra. Relatively, the spectra agree to within
4\%, with the depression being more pronounced in the
5\arcsec-calibrated spectra, i.e. there is no dependence on wavelength
(e.g. due to colour-differential slit losses) beyond the 4\% level.
We note that Fig.~\ref{fig:spectrum991216} shows the
1\arcsec-calibrated spectrum. Moreover, we have also performed the
flux calibration using observations of three different 5\arcsec\
standards taken during three different photometric nights, up to one
week before the night of the GRB. The sensitivity curves of all these
standards differ by at most 8\% in absolute terms (wavelength range:
4000--6000~\AA) with the 5\arcsec-standard taken during the
\9\ night. Therefore, we regard it unlikely that the blue flux
depression in the spectrum of \9\ is caused by an error in the flux calibration.

Assuming that the relative flux calibration is acceptable, and that
the redshift of \9\ is $z=1.02$, we investigate whether the depression
in the blue part of the spectrum can be due to a redshifted 2175~\AA\
absorption feature, as observed in the Galactic extinction curve. With
this aim, we fit the Milky Way (MW) extinction curve of
\citet{1992ApJ...395..130P} to the regions of our spectrum
void of identified absorption lines, and fixing the redshift at
$z=1.02$.  We assume that the intrinsic afterglow spectrum is a single
power law, to obtain the intrinsic spectral slope ($\beta$) and
host-galaxy extinction ($A_V$) that best matches the observed spectrum
(i.e. minimum $\chi^2$). For the MW we obtain the following best-fit
values: $\beta=-0.06\pm0.02$, corresponding to a rest-frame V-band
extinction of $A_V=0.16\pm0.02$~mag. The resulting fit is shown by the
smooth solid line in Fig.~\ref{fig:spectrum991216}. Fitting the red
part (6000-9300~\AA) of the \9\ spectrum with a single power law model
results in a slope of $\beta=-0.187\pm0.014$. We have also fit the
Large and Small Magellanic Clouds' (LMC and SMC) extinction curves;
these are shown by the dashed (LMC) and dotted (SMC) curves in
Fig.~\ref{fig:spectrum991216}. These MC fits are less successful due
to the 2175~\AA\ feature being less prominent in the LMC as compared
to the MW, and absent in the SMC.  Although the MW fit is reasonable,
the 2175~\AA\ feature clearly does not correspond to the minimum
depression in the spectrum. Placing the burst at a redshift of
$z=1.19$, or alternatively, moving the peak of the bump from 2175~\AA\
to 2360~\AA\, results in the best fit. This fit is shown by the
dash-dotted line in Fig.~\ref{fig:spectrum991216}.

Although it seems likely that some form of graphitic carbon (possibly
polycyclic aromatic hydrocarbon -- PAH -- molecules) is responsible
for the 2175~\AA\ absorption feature in our Galaxy
\citep{2003ARA&A..41..241D}, its exact nature is still unclear after
decades of research. Observations along different Galactic sight lines
have shown that although the FWHM of the 2175~\AA\ bump can vary
considerably (in our \9\ fits we did not allow this width to change),
the central wavelength of the bump is very stable
\citep[$2174\pm10$~\AA;][]{1986ApJ...307..286F,2003ARA&A..41..241D}.
However, in a few particular cases where the UV component of the star
used to measure the interstellar extinction is strong enough to reveal
extinction by grains in its own circumstellar material, a broad bump
is observed with a central wavelength in between 2300~\AA\ and
2500~\AA. Abell~30 \citep{1981ApJ...245..124G} and HD~213985
\citep{1989ApJ...347..977B} are two examples of such UV-strong,
hydrogen-poor stars. It is tempting to make the connection with the
UV-strong environment of a GRB, which may cause the extinction bump to
be redder than 2175~\AA\ as well.  Also, laboratory experiments
\citep[see][]{1991ApJ...382L..97B,1996ApJ...462.1020B}, aimed at
testing the physical grain models proposed by
\citet{1986ApJ...305..817H} and \citet{1990MNRAS.243..570S},
have shown that, by decreasing the hydrogen content in carbon grains,
the 2175~\AA\ bump not only becomes more pronounced, but also shifts
to higher wavelengths, up to 2600~\AA\ \citep{1995ApJ...444..288M}.
Beyond the Local Group, the best case for detection of the Galactic
extinction feature is a $z=0.83$ gravitational lens system, where the
best-fit central wavelength of the bump is found to be only slightly
larger than the Galactic value: 2234$\pm$24~\AA\
\citep{2002ApJ...574..719M}. Therefore, the flux depression in
our spectrum may have been caused by such a red extinction
bump along the line-of-sight to \9. 

\subsection{HST imaging}

\begin{figure}
  \centering \includegraphics[width=8.5cm]{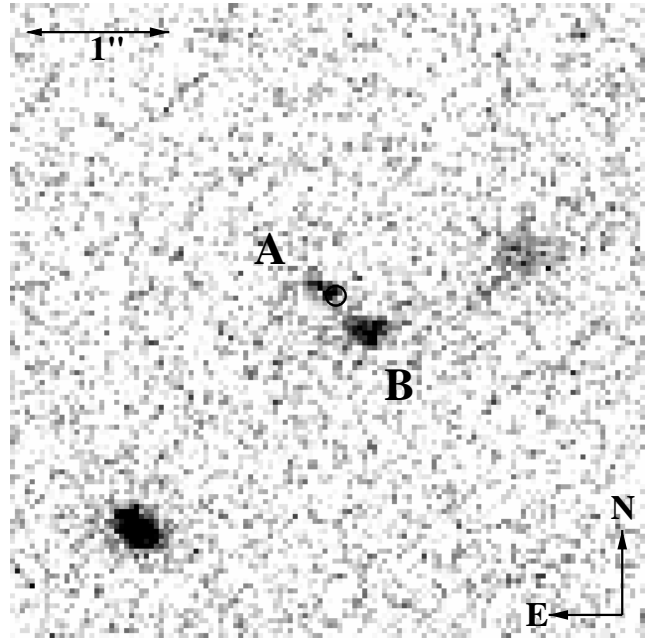} \caption{The
  sum of the HST/STIS 50CCD and LP images of the field of \9. The
  position of the early optical transient is marked with a black
  circle, with an error radius of 0\farcs05 (2 drizzled HST pixels).
  The regions A and B could be part of the same galaxy at $z=1.02$, or
  two systems in the process of merging. Based on the identification
  of at least two absorption-line systems in the spectrum, an
  alternative configuration is that A is the host galaxy of \9\ at
  $z=1.02$, and that B is responsible for the absorption lines at
  $z=0.80$. Either one of the galaxies that are located 2\arcsec\ to
  the SE and NW of the transient position could be the counterpart of
  the tentative $z=0.77$ system.  \label{fig:image991216}}
\end{figure}

The field of GRB 991216 was observed with the Space Telescope Imaging
Spectograph (STIS) approximately 4 months after the burst, on 17 April
2000, starting at 11:36 UT, through the clear (50CCD) and long pass
(LP) filters, each for a total of 4790 seconds.  The pipeline reduced
images were drizzled \citep[see][]{2002PASP..114..144F} onto output
images with pixels one-half native scale, or approximately 0\farcs025
on a side. Figure~\ref{fig:image991216} shows the central 5$\times$5
square arcseconds of the sum of the 50CCD and LP images.

We have projected the optical afterglow (OA) position from an early
VLT image, taken 1.5 days after the burst, to the frame of the
HST drizzled images. Four bright nearby reference stars were
used, and the estimated 1$\sigma$ error in the resulting position is
0\farcs05, corresponding to 2 drizzled pixels.  The position and its
error are indicated with a circle in Fig.~\ref{fig:image991216}. The
error circle coincides with one of two faint regions of light (A and
B), which are separated by 0\farcs6. The two regions are about equally
bright in the 50CCD as in the LP image, showing that they have similar
colours with most of the light coming out longward of 5500~\AA. A and
B are possibly part of the same galaxy, or they could be two systems
that are interacting. Several other host galaxies also show a complex
morphology, e.g. the hosts of GRBs 980613, 990123, 000926, and
011211. The likely detection of the Ca{\sc ii} doublet at $z=1.02$
argues slightly in favour of this merger scenario, since (at low
redshifts) the Ca{\sc ii} lines are detected mainly in disrupted
environments \citep{1991MNRAS.251..649B,1992ApJ...399..373C}. Another
possibility is that B is actually a foreground system at $z=0.80$,
corresponding to the absorption-line system detected in the VLT
spectrum, and that A is the host galaxy of \9\ at $z=1.02$. The
detection of emission lines from A and B would determine which of
these possibilities is correct. Either one of the two other galaxies
that are located roughly 2\arcsec\ away from the OA position could be
responsible for the tentative absorption system at $z=0.77$.  At a
redshift of 0.77, 2\arcsec\ corresponds to about 14 kiloparsec
(assuming $H_0=65$~km s$^{-1}$ Mpc$^{-1}$, $\Omega _{\rm m}=0.3$, and
$\Omega_{\Lambda}=0.7$), which is smaller than the typical galaxy halo
size, and so one would indeed expect to detect Mg{\sc ii} in
absorption.

Using an aperture radius of 0\farcs2, we measured the following fluxes
in photons per second in the drizzled 50CCD and LP images:
$0.36\pm0.03$ for A in 50CCD, $0.71\pm0.03$ (B, 50CCD), $0.33\pm0.02$
(A, LP), and $0.45\pm0.02$ (B, LP). We also measured the total flux of
A and B combined, with an aperture radius of 1\farcs1: $1.90\pm0.15$
(A+B, 50CCD), and $1.23\pm0.13$ (A+B, LP). This aperture size is the
same as used by \citet{2000ApJ...543..697H}, who obtained
$R_C=24.8\pm0.2$ for the host galaxy from a Keck II image taken on 4
April 2000, i.e. about two weeks before the HST imaging reported
here. To convert the HST fluxes to standard magnitudes, a spectral
energy distribution has to be assumed for the source. Using {\it
synphot} within IRAF, combined with the Kinney-Calzetti atlas of
galaxy spectra \citep{1994ApJ...429..582C,1996ApJ...467...38K}, we
found that the best-fitting galaxy template is the Sc template
(redshifted to $z=1.02$). The redshifted starburst templates also
provide reasonably consistent magnitudes for both passbands and a
total magnitude corresponding with the value of $R_{\rm C}=24.8\pm0.2$
of Halpern and colleagues. Adopting the Sc galaxy template, the fluxes
above can be converted to the following magnitudes: $R_{\rm C}(\rm
A)=26.5$, $R_{\rm C}(\rm B)=26.0$ and $R_{\rm C}(\rm A+B)=24.9$, and
colours: $V-R_{\rm C}(\rm A)=1.1$, $V-R_{\rm C}(\rm B)=1.0$, and
$V-R_{\rm C}(\rm A+B)=1.0$. We estimate the uncertainty in the
transformation of fluxes to magnitudes to be of the order of 0.2 mag.
We note that these magnitudes (including the one of Halpern et al.)
and colours have not been corrected for the Galactic extinction along
the \9\ sight line ($A_V$=2.1 and $A_R$=1.7).

The transient afterglow may still be present in these observations,
but the low S/N does not allow an unambiguous identification of the
bright patch at the edge of the galaxy as a point source.  We estimate
that any remaining OA is not brighter than $R=27.6$. Assuming the
single power law decay index, $\alpha=-1.36$, of
\citet{2000ApJ...543...61G}, the expected magnitude of the afterglow
at the time of our observations is $R\sim27$. Our observations
therefore tend to confirm the break in the light curve reported by
\citet{2000ApJ...543..697H}. A supernova as bright as SN1998bw at a
redshift of $z=1.02$ would have $R>30$ at the epoch of our
observations, and would thus be too faint to be detected.

\section{GRB~011211}
\label{sec:grb011211}

Following the {\it BeppoSAX} localisation of \1\
\citep{2001GCN..1188....1G,2002GCN..1215....1F}, the afterglow was
discovered by \citet{2001GCN..1191....1G} \citep[see
also][]{2004NewA....9..435J}, and a preliminary redshift of $z=2.14$
was reported by \citet{2001GCN..1200....1F}, which was confirmed by
\citet{2001GCN..1209....1G}. Rapid optical variability (1 hour
variability time scale at 12 hours after the burst) is detected in the
afterglow lightcurve; these variations could be caused by either
density fluctuations in the external medium
\citep{2002AJ....124..639H,2004NewA....9..435J}, or by a non-uniform
jet structure \citep{2004NewA....9..435J}.
\citet{2002Natur.416..512R} obtained an XMM-Newton 
spectrum of the X-ray afterglow and detect transient features that
they identify as \ion{Mg}{xi}, \ion{Si}{xiv},
\ion{S}{xvi}, \ion{Ar}{xviii}, and \ion{Ca}{xx} at a redshift of 
$z=1.88\pm0.06$, i.e. blue-shifted by about $0.1c$ with respect to the
optical redshift.  The authors suggest this to be due to reprocessing
of the burst flux in a shell expelled by a supernova explosion
preceding the GRB by 4 days. However, \citet{2003MNRAS.339..600R}
\citep[but see][]{2003A&A...403..463R} and \citet{sako} claim that
the detection of these metal lines in the X-ray spectrum is not
significant. The host galaxy has $R=24.95\pm0.11$, with a modest
($A_V\sim0.1$) SMC-like extinction, and the most likely afterglow
model is that of a jet expanding into a constant mean density
environment \citep{2003A&A...408..941J}. \citet{fynbolya} detect \lya\
in emission in the \1\ host galaxy, extending up to 1\arcsec\ to the
North-North-East of the afterglow position.

\begin{figure*}[t]
  \centering \includegraphics[width=15cm]{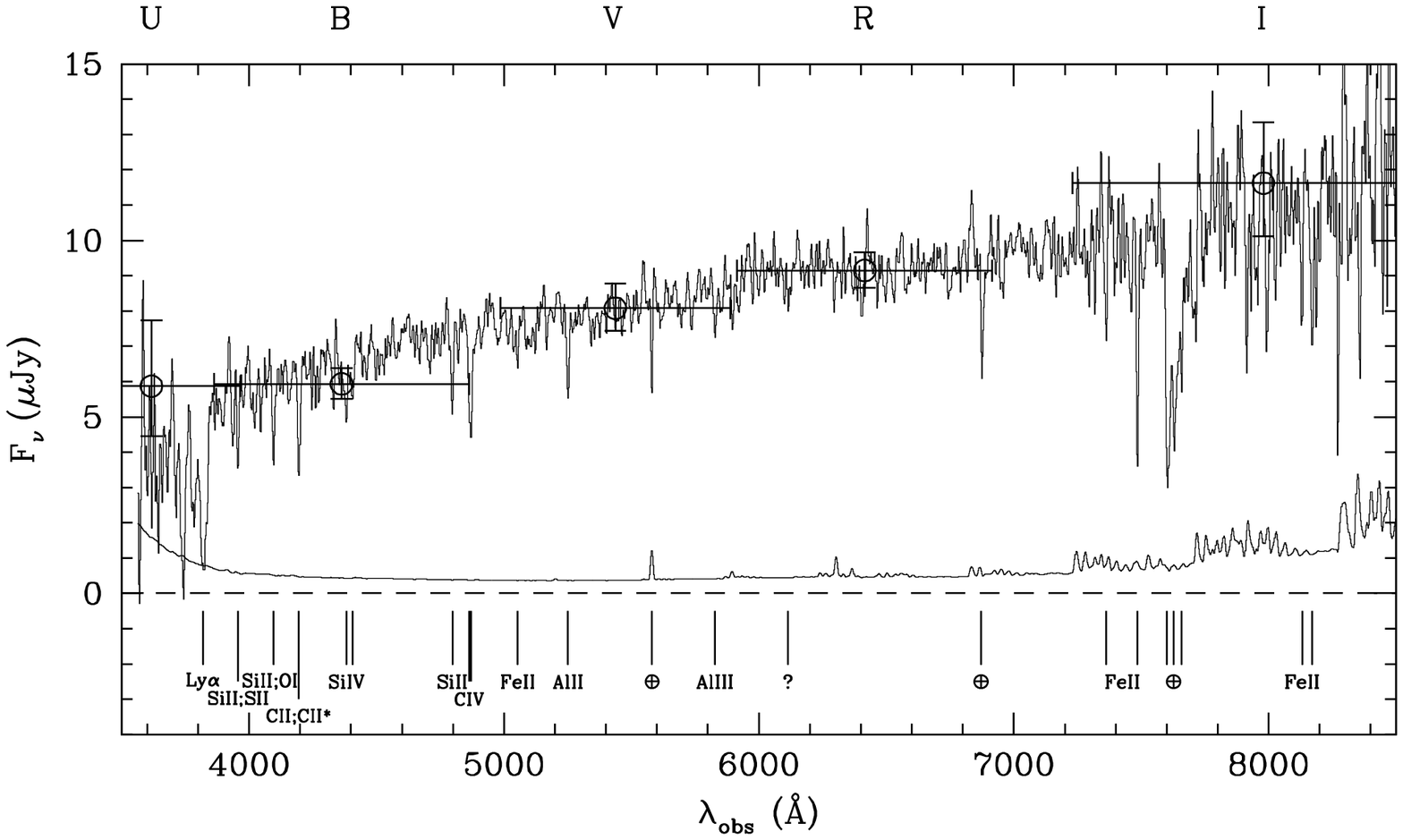}
  \caption{The optical spectrum of the afterglow of \1, smoothed with
  a boxcar the size of 3 pixels, or about 8~\AA.  The resolution is
  approximately 10~\AA\ across the entire spectrum, which corresponds
  to a resolving power of 600 at 6000~\AA. The spectrum flux level has
  been scaled downward by 70\% to match the photometric $UBVRI$ data
  points of \citet{2003A&A...408..941J}, which are shown by the open
  circles.  With one exception, all significant (3$\sigma$) features
  (see Table~\ref{tab:lines011211}) can be identified with \lya\ and
  metal-absorption lines in a single system, the host galaxy, at a
  redshift of $z=2.1418\pm0.0018$. The lack of \lya-forest lines
  redward of \lya\ shows that \1\ occurred at this redshift.  The
  bottom solid line shows the 1$\sigma$ Poisson error.
  \label{fig:spectrum011211}}
\end{figure*}

\subsection{Absorption lines and redshift}
\label{sec:011211spectroscopy}

Figure~\ref{fig:spectrum011211} shows the spectrum of \1.  As was
discussed in Sect.~\ref{sec:991216shape}, the night in which the
\1\ afterglow spectroscopy was performed was not photometric. Comparison
with the photometric calibration of \citet{2003A&A...408..941J} shows
our calibration to be off by 70\%, and we have scaled the spectrum
down by this amount in Fig.~\ref{fig:spectrum011211}. The photometric
observations of \citet{2003A&A...408..941J}, scaled in time to the
epoch of the spectrum, are shown by the open circles in
Fig.~\ref{fig:spectrum011211}. The horizontal error bars reflect the
FWHM of the filter transmission. Relatively, the spectral and
photometric flux calibration match reasonably well. Fitting a single
power law to our spectrum over the range 4000--8000~\AA\ results in a
spectral slope of $\beta=-0.86\pm0.02$.

Table~\ref{tab:lines011211} lists the lines detected with a
significance above 3$\sigma$ in the spectrum. The redshift as
determined from the detected low-ionisation metal lines is:
$z=2.1418\pm0.0018$. \citet{2002AJ....124..639H} have reported the
detection of 8 metal absorption lines at a mean redshift of
$z=2.140\pm0.001$, in a spectrum taken with the LDSS-2 imaging
spectrograph at the Magellan 6.5m telescope.  Although the S/N of our
spectra is superior, we are able to confirm the detection of only 4 of
the lines that \citet{2002AJ....124..639H} identify. Their
\ion{Si}{iv}~\l 1368, \ion{Cr}{ii}~\l\l 1406, 1422 are 
misidentifications, and the line at 6875.2~\AA\, identified as
\ion{Fe}{iii}~\l 2189 by Holland and colleagues, is a well-known
atmospheric absorption feature. Also, we clearly do not detect the
broad absorption feature that they detect at 4600~\AA.

\begin{table}[tbp]
  \centering
  \caption[]{Lines detected above 3$\sigma$ in the spectrum of
\1. \label{tab:lines011211}}
  \null\vspace{-1.0cm}
  $$
  \begin{array}{crrc}
    \hline
    \hline
    \noalign{\smallskip}            
    \lambda_{\rm obs} &
    W_{\rm obs} (\rm\AA) &
    \rm ID &
    z_{\rm abs} \\
    \hline
3820.0  &  16.6  \pm  2.1  & $\lya$       \,\lambda 1215             &  2.1423 \\
3957.0  &   4.7  \pm  1.1  & \ion{Si}{ii}\,\lambda 1260              &  2.1394 \\
        & \rm blended\,\,with & \ion{S}{ii}\,\lambda 1259	     &\\
4096.9  &   4.6  \pm  0.7  & \ion{Si}{ii}\,\lambda 1304              &  2.1409 \\
        & \rm blended\,\,with & \ion{O}{i}\,\lambda 1302             &\\
4195.6  &   5.6  \pm  0.8  & \ion{C}{ii}\,\lambda 1334               &  2.1439 \\
        & \rm blended\,\,with & \ion{C}{ii}^*\,\lambda 1335          &\\
4381.7  &   2.9  \pm  0.6  & \ion{Si}{iv}\,\lambda 1393              &  2.1438 \\
4405.9  &   2.0  \pm  0.6  & \ion{Si}{iv}\,\lambda 1402              &  2.1409 \\
4797.8  &   3.6  \pm  0.5  & \ion{Si}{ii}\,\lambda 1526              &  2.1426 \\
4863.4  &   2.2  \pm  0.4  & \ion{C}{iv}\,\lambda 1548               &  2.1413 \\
4870.9  &   3.7  \pm  0.5  & \ion{C}{iv}\,\lambda 1550               &  2.1410 \\
5053.7  &   1.7  \pm  0.5  & \ion{Fe}{ii}\,\lambda 1608              &  2.1420 \\
5251.7  &   3.5  \pm  0.5  & \ion{Al}{ii}\,\lambda 1670              &  2.1432 \\
5828.3  &   1.7  \pm  0.4  & \ion{Al}{iii}\,\lambda 1854             &  2.1424 \\
6114.2  &   1.3  \pm  0.4  &                                         &\\
7362.8	&   3.1  \pm  1.0  & \ion{Fe}{ii}\,\lambda 2344              &  2.1408 \\
7485.5  &   6.5  \pm  0.8  & \ion{Fe}{ii}\,\lambda 2382              &  2.1415 \\
8131.4  &   3.8  \pm  1.0  & \ion{Fe}{ii}\,\lambda 2586              &  2.1436 \\
8171.4  &   3.5  \pm  1.0  & \ion{Fe}{ii}\,\lambda 2600              &  2.1426 \\
    \hline
  \end{array}
  $$
%  \begin{list}{}{} 
%  \end{list}
\end{table}

\subsection{The \ion{H}{i} column density, and metallicity}

\begin{figure}[t]
  \centering \includegraphics[width=8.5cm]{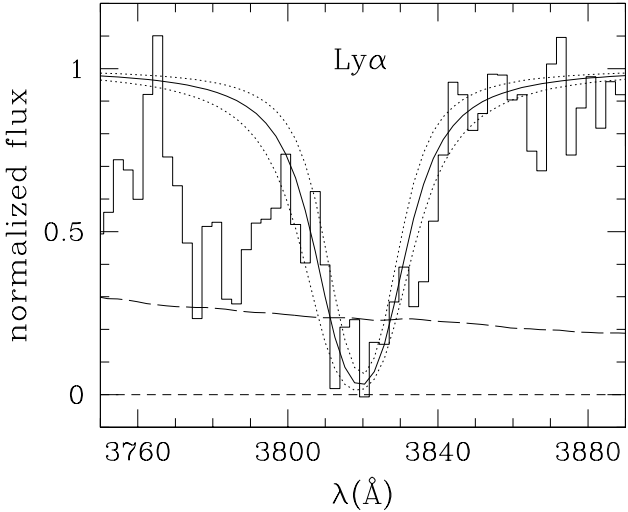} \caption{The
  fit (solid continuous line) of the \lya\ line to the normalised
  spectrum of \1\ (solid histogram), resulting in
  log~$N$(\ion{H}{i})=$20.4\pm0.2$.  The dotted lines show the
  1$\sigma$ errors of the fit, and the long-dashed line indicates the
  1$\sigma$ Poisson error spectrum.  \label{fig:lya011211}}
\end{figure}

Figure~\ref{fig:lya011211} shows the fit to the \lya\ line in the
spectrum of \1, performed within the LYMAN context in Midas. The
resulting neutral hydrogen \ion{H}{i} column density is
log~$N$(\ion{H}{i})=$20.4\pm0.2$, with the line center at
($3819.7\pm2.1$)~\AA, corresponding to a redshift of
$z=2.142\pm0.002$. The fit region is limited to around the core of the
line, 3810--3833~\AA. If this is extended to the red, the continuum
around 3850~\AA\ is very poorly fit, and to the blue of 3810~\AA\ the
\lya\ forest starts to contaminate the \lya\ line.

The \ion{H}{i} column density in the host of \1\ is at the low end of
the \ion{H}{i} GRB afterglow column densities obtained so far. Of the
7 afterglows to date for which it was possible to detect \lya\ in
absorption at the host-galaxy redshift, 6 turn out to be damped \lya\
systems \citep[see][]{2001A&A...370..909J,hjorth020124,vrees030323}
and therefore have log~N(\ion{H}{i}) $>$ 20.3. The host of GRB\,021004
is the exception, with a column density $N$(\ion{H}{i}) $\sim
5\times10^{19}$ atoms cm$^{-2}$.

The low resolution of our spectrum makes it difficult to fit an
absorption profile to the detected metal lines to obtain the
corresponding column density and metallicity. However, we can derive a
strict lower limit on these quantities by simply assuming that the
lines are not saturated, even though most lines are in fact
saturated. In this optically thin approximation there is a linear
correspondence between the equivalent width of a line, and its column
density: log~$(W_{\rm rest}/\lambda)= $log~$(Nf\lambda) - 20.053$,
where $f$ is the oscillator strength
\citep[see][]{2003ApJS..149..205M}, and the unit of $\lambda$ is
\AA. Applying this relation to the detected low-ionisation lines that
are not blended, we obtain: log~$N(\ion{Si}{ii}) \ge 14.6$,
log~$N(\ion{Fe}{ii}) \ge 14.6$, and log~$N(\ion{Al}{ii}) \ge
13.4$. Assuming that these ions are the dominant ionisation state for
the corresponding element, these column density limits can be
translated to the following metallicity limits: [Si/H]$\ge -1.3$,
[Fe/H]$\ge -1.3$, and [Al/H]$\ge -1.5$, where we have used the solar
abundances from \citet{1998SSRv...85..161G} \citep[see
also][]{2003ApJS..149..205M}, and the \ion{H}{i} column density
derived above. Neither \ion{Zn}{ii}~\l 2026 nor \ion{Zn}{ii}~\l 2062
are detected, for which we measure: $W_{\rm
obs}(\ion{Zn}{ii}~2026)=0.7\pm0.7$~\AA\ and $W_{\rm
obs}(\ion{Zn}{ii}~2062) = -0.1\pm0.5$~\AA. These values correspond to
the 2$\sigma$ column density limit: log~$N(\ion{Zn}{ii})<13.4$ and the
metallicity limit: [Zn/H]$<+0.3$.

\begin{figure}[b]
  \centering \includegraphics[width=8.5cm]{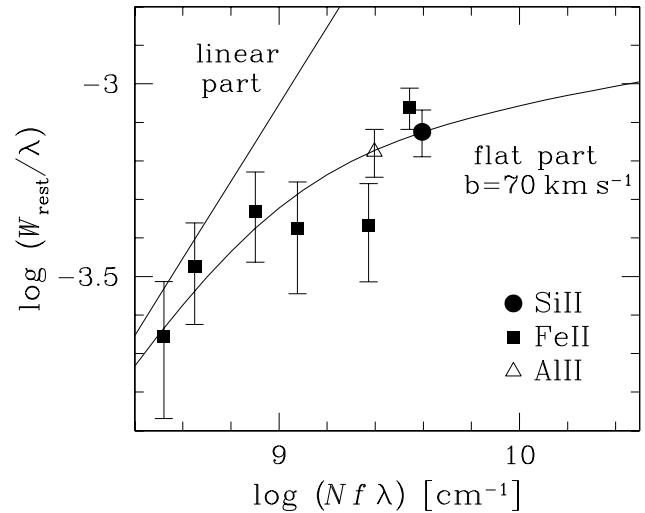} \caption{The curve
  of growth for the absorption lines measured in the afterglow of \1.
\label{fig:cog}}
\end{figure}

Taking this a step further, the abundances can be estimated through
the curve-of-growth (COG) technique
\citep[see][]{1978ppim.book.....S}. Our COG analysis is very similar 
to the one that \citet{2003ApJ...585..638S} applied to three GRB host
galaxies.  Figure~\ref{fig:cog} shows the resulting curve of growth
for the absorption lines measured in the afterglow of \1. The straight
line corresponds to the linear part, where the lines are optically
thin. The bent curve shows the deviation from the optically thin
regime due to saturation of the absorption lines, the so-called flat
part. The degree of bending depends on the amount of Doppler
broadening ($b$ in \kms) in the medium in which the lines
originate. The best fit curve is obtained by varying the column
density of each element separately (i.e. sliding the data points
horizontally), and finding the value for the Doppler width $b$ that
fits best with the overall shape of all elements together.  We have
assumed a Gaussian shape for the line profiles, i.e. we have neglected
damping or natural line broadening, which is important for high column
density absorbers. We only use the low-ionisation lines from
Table~\ref{tab:lines011211} which are not obviously blended, i.e.
\ion{Si}{ii}~\l\l 1526, \ion{Fe}{ii}~\l\l 1608, 2344,
2382, 2586, 2600, and \ion{Al}{ii}~\l 1670. We also included the
measurement of \ion{Fe}{ii}~\l 2374, even though this line is detected
at the 2.5$\sigma$ level only; for this reason it is not listed in
Table~\ref{tab:lines011211}. Minimising $\chi^2$, we find the
following values for the Doppler width ($b$) and the
\ion{Si}{ii}, \ion{Fe}{ii} and \ion{Al}{ii} column densities:
$b$=\gpm{70}{20}{10}~\kms, log~$N$(\ion{Si}{ii})=\gpm{15.1}{0.6}{0.3},
log~$N$(\ion{Fe}{ii})=$14.7\pm0.2$, and
log~$N$(\ion{Al}{ii})=\gpm{13.9}{0.5}{0.3}. The errors in these values
are obtained by varying each parameter (while fitting the others), up
to the point where the chi-squared becomes the minimum chi-squared
plus one, i.e. where $\chi^2$=$\chi^2_{\rm min}$+1. The full range
that the parameter can adopt within this chi-squared restriction is
taken as the 1$\sigma$ error range. These metal column densities
correspond to the following metallicities:
[Si/H]=\gpm{-0.9}{0.6}{0.4}, [Fe/H]=$-1.3\pm0.3$, and
[Al/H]=\gpm{-1.0}{0.5}{0.3}.  Again, we have made the common
assumption that these low-ionisation species are the dominant
ionisation states for the corresponding element.

Finally, we do not see the \lya\ in emission that was detected by
\citet{fynbolya}, either in the 2-dimensional spectra or in the
extracted spectra. This is due to the poor S/N of our spectrum around
3800~\AA.

\section{GRB~021211}
\label{sec:grb021211}

The {\it High Energy Transient Explorer 2} (HETE-2) satellite detected
\2\ (HETE-2 trigger \#2493) at 11:18:34 UT on 2002 December 11, as a
bright and X-ray-rich burst \citep{2003ApJ...599..387C}.  The optical
afterglow was imaged as early as 65 seconds after the GRB
\citep{2002GCN..1757....1W}, and was reported by
\citet{2003ApJ...586L...5F} to the GRB community within an hour of the
burst. The afterglow was fainter than nearly all known afterglows to
that date at an epoch of 1 day after the GRB \citep[see Fig.~2
of][]{2003ApJ...586L...5F}. \citet{2003ApJ...586L...9L} find evidence
for a light-curve break at $t\sim10$ minutes, with the decay index
becoming more shallow from --1.82 to --0.82, which they suggest is due
to the emission before the break being dominated by the reverse shock.
\citet{2004MNRAS.353..511P} calculate the shock microphysical
parameters, and find that the reverse-forward shock scenario provides
a more natural explanation than the wind-bubble scenario for the steep
early decay. Finally, \citet{2003A&A...406L..33D} present
spectroscopic evidence for a supernova component in the late-time
afterglow of \2.

\subsection{Emission line and redshift}

Our team first reported a redshift of $z=0.800$ for the putative host
galaxy of \2, based on the identification of the [\ion{O}{ii}] and
[\ion{O}{iii}] emission lines \citep{2002GCN..1756....1V}. However, we
later found that, due to saturation of the offset star, the slit
position did in fact not cover the afterglow position
\citep{2002GCN..1767....1V}.  Instead, the slit mostly contained the
galaxy 1\farcs5 to the North-East of the afterglow, first noted by
\citet{2002GCN..1750....1M}, whose redshift is then $z=0.800$,
and which is unrelated to the GRB as shown below.

\begin{figure*}[t]
  \centering \includegraphics[width=15cm]{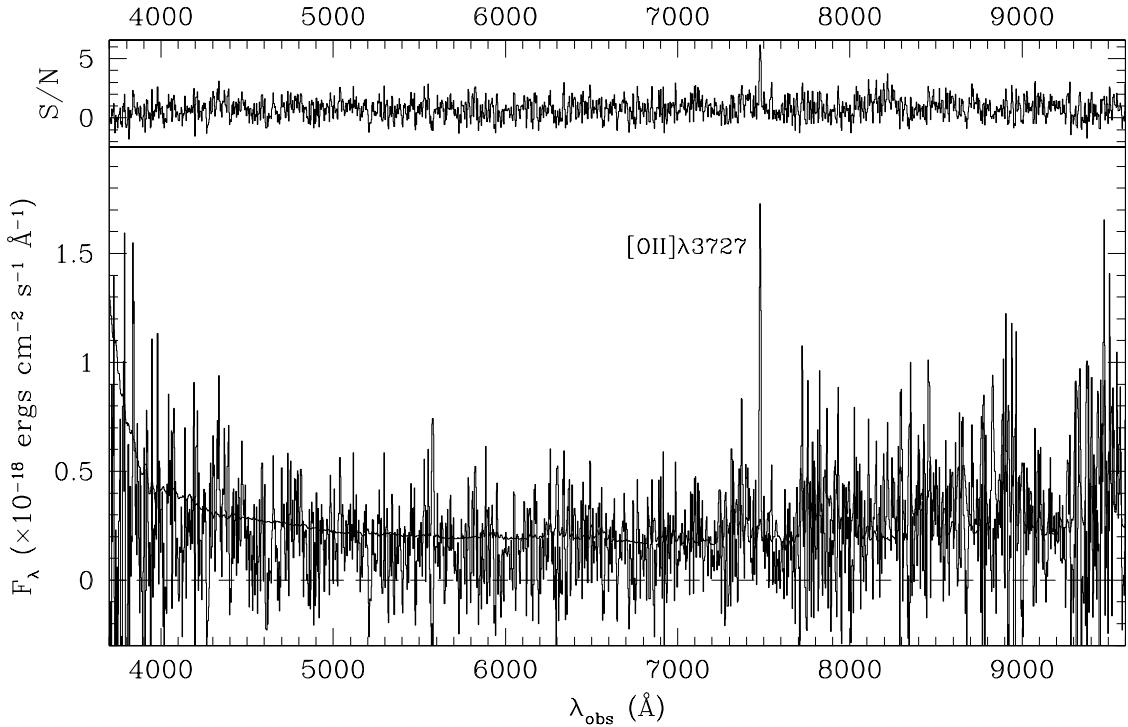}
  \caption{The optical spectrum of the probable host galaxy of \2. The
  spectrum has neither been flux calibrated nor smoothed.  The
  resolution is approximately 11~\AA\ across the entire spectrum,
  which corresponds to a resolving power of 550 at 6000~\AA.  The top
  panel shows the S/N spectrum, smoothed with a boxcar with a size
  similar to the spectral resolution.  Only 1 significant line is
  detected, at 7478~\AA, which is most likely [\ion{O}{ii}]~\l 3727 at
  a redshift of $z=1.006$. The 1$\sigma$ Poisson error spectrum is
  also plotted in the bottom panel. \label{fig:spectrum021211}}
\end{figure*}

\begin{figure*}[t]
  \centering \includegraphics[width=15cm]{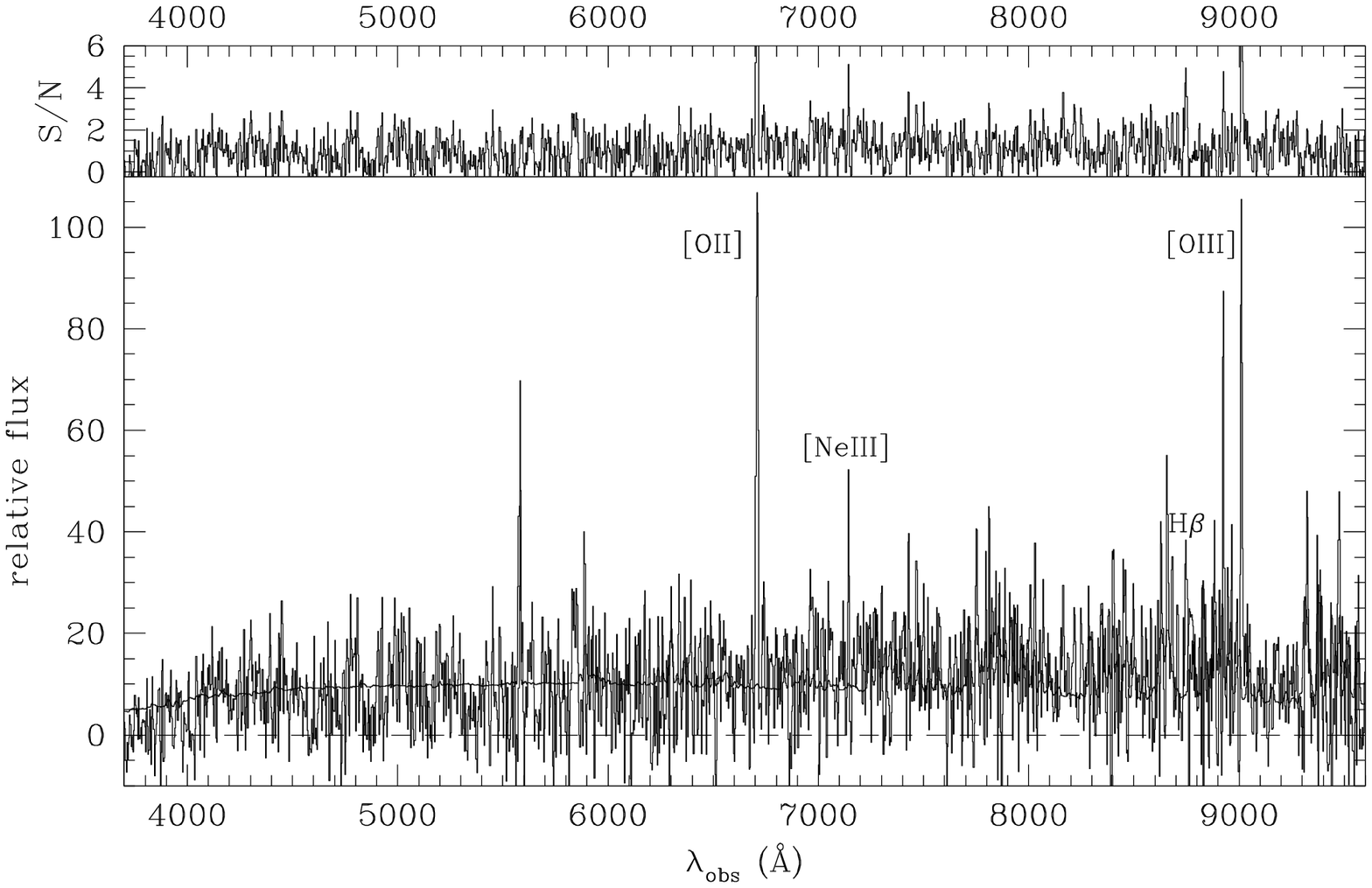}
  \caption{The optical spectrum of the galaxy 1\farcs5 to the
  North-East of the \2\ afterglow location.  The resolution is
  approximately 11~\AA\ across the entire spectrum, which corresponds
  to a resolving power of 550 at 6000~\AA.  The spectrum has neither
  been flux calibrated nor smoothed.  The top panel shows the S/N per
  pixel spectrum.  The significant lines can be identified with
  [\ion{O}{ii}]~\l 3727, [\ion{Ne}{iii}]~\l 3968, H$\beta$ and
  [\ion{O}{iii}]~\l\l 4959, 5007 at a redshift of $z=0.800$. The
  1$\sigma$ Poisson error spectrum is also plotted in the bottom
  panel.  \label{fig:galaxy021211}}
\end{figure*}

In follow-up spectroscopy of \2, we positioned the slit to cover both
the position of the early afterglow and that of the galaxy at
$z=0.800$, as well as a nearby reference star. Using this reference
star and the slit position angle, the exact location of both the
galaxy and the projected position of the early-time afterglow on the
2-dimensional spectrum can be inferred.
Figure~\ref{fig:spectrum021211} shows the spectrum extracted at the
position of the early afterglow; we find one clear emission line at
7478~\AA. This line most likely corresponds to [\ion{O}{ii}]~\l 3727
emission from the host galaxy of \2\ at a redshift of $z=1.006$
\citep{2002GCN..1767....1V}.  This redshift was later confirmed by
\citet{2003GCN..1809....1V} who besides [\ion{O}{ii}] also detect
H$\beta$~\l 4861 and [\ion{O}{iii}]~\l\l 4959, 5007. The limited
spectral coverage in the red part of our spectrum does not allow us to
detect these lines. For completeness, we show in
Fig.~\ref{fig:galaxy021211} the wavelength-calibrated spectrum of the
galaxy that is located 1\farcs5 to the North-East of the afterglow,
with a redshift of $z=0.800$.

\citet{2003A&A...406L..33D} obtained a total of 8 hours of
spectroscopy with the VLT/FORS2 and the 150I grism and OG590 filter
combination, which covers the optical spectrum above 6000~\AA\ with a
resolution of 19~\AA. \citet{2003A&A...406L..33D} focus on the
possible evidence for a supernova component in the late-time afterglow
of \2, and briefly mention the detection of H$\beta$~\l 4861 and
[\ion{O}{iii}]~\l\l 4959, 5007.  Since the host-galaxy metallicity can
be measured from the combination of [\ion{O}{ii}], H$\beta$~\l 4861
and [\ion{O}{iii}]~\l\l 4959, 5007 \citep[see][]{1999ApJ...514..544K},
we decided to reduce and extract these archival spectra as well.
However, we are unable to detect any significant line apart from the
line at 7478~\AA\ mentioned above.  This is probably due to a
combination of the very low spectral resolution, which complicates the
subtraction of the numerous sky emission lines in the red, and the low
sensitivity around 10,000~\AA, where the [\ion{O}{iii}] and H$\beta$
lines are expected.

Figure~\ref{fig:spectrum021211} shows that the continuum level just
redward of the significant line at 7478~\AA\ is similar to that on the
blue side, suggesting that the line is not
\lya\ at $z=4.15$. The most likely identification is [\ion{O}{ii}]~\l
3727 at a redshift of $z=1.006$. This line is actually a doublet with
components at 3727.09~\AA\ and 3729.88~\AA\ (vacuum wavelengths), but
our spectral resolution is too low to resolve these.  At a redshift of
$z=1.006$, other prominent emission lines such as H$\beta$~\l 4861,
and [\ion{O}{iii}]~\l\l 4959, 5007 are not covered by our spectral
range. If the significant line were H$\alpha$ at $z=0.139$, we would
have expected to detect [OII] (at 4246~\AA) and the [OIII] lines (at
5649~\AA\ and 5705~\AA) in the blue part of the spectrum, for which we
do not find any evidence. And finally, if the line were
[\ion{O}{iii}]~\l 5007 at $z=0.49$, we would have expected to detect
[\ion{O}{iii}]~\l 4959 (at 7406~\AA) with roughly one-third of the
strength of [\ion{O}{iii}]~\l 5007, and [\ion{O}{ii}] at 5566~\AA,
even though this latter detection would have been made slightly
difficult due to the presence of a strong sky line at 5577~\AA. The
[\ion{O}{ii}]~\l 3727 at $z=1.006$ identification is strengthened by
the observed drop between the 8000--9000~\AA\ continuum and that
blueward of the emission line. This is consistent with the red HST
$V-I$ colour reported by \citet{2002GCN..1781....1F}, and with the
possibility that this colour is caused by the 4000~\AA\ break.

\subsection{Star-formation rate}

Assuming that the detected emission line indeed is [\ion{O}{ii}], we
can estimate the corresponding star-formation rate using the relation
$SFR_{\rm [OII]} = 1.4 \times 10^{-41} L_{\rm [OII]}$ M\subsun\
yr$^{-1}$ from \citet{1998ARA&A..36..189K}, where $L_{\rm [OII]}$ is
in erg s$^{-1}$. We measure an [OII] flux of
($1.92\pm0.15$)$\times10^{-17}$~erg s$^{-1}$ cm$^{-2}$, where the
error is dominated by the uncertainty in the continuum level. As the
continuum level is close to zero, the measurement of the observed
equivalent width is highly uncertain: $W_{\rm [OII]\,obs}$ =
\gpm{90}{110}{40}~\AA. At the luminosity distance of \2\ of
$2.06\times10^{28}$~cm, the [\ion{O}{ii}] flux corresponds to a
star-formation rate of $SFR_{\rm [OII]}=1.4$~M\subsun\ yr$^{-1}$,
which is a typical value of GRB host galaxies \citep[ranging from 0.3
to 24 M\subsun\ yr$^{-1}$, e.g. see Table~1 of][]{vrees1}. We note
that this value has not been corrected for host-galaxy extinction, and
is therefore a strict lower limit to the actual star-formation rate.

\section{Conclusions}
\label{sec:conclusions}

We have determined the redshift for three GRBs: $z=1.02$ for \9,
$z=2.142$ for \1 and $z=1.006$ for \2. For \9, we also find evidence
for a foreground absorption system at $z=0.80$, and possibly another
at $z=0.77$. We argue that the flux depression from 4000--5500~\AA\ in
the \9\ spectrum is unlikely to be due to an error in the flux
calibration, and find that it could be explained by a host-galaxy
extinction bump similar to the Galactic 2175~\AA\ feature, but at a
wavelength of 2360~\AA. In the Galaxy, similar extinction bumps, with
a central wavelength between 2300~\AA\ and 2500\AA, have been observed
toward UV-strong, hydrogen-poor stars. Fitting the \lya\ absorption
line that we detect in the afterglow spectrum of \1, we obtain a
neutral hydrogen column density of log~$N$(\ion{H}{i})=$20.4\pm0.2$,
indicating that it is a damped \lya\ (DLA) system. This is at the low
end of the \ion{H}{i} column densities detected in GRB
afterglows. Using a curve-of-growth analysis, we find the following
metallicity estimates at the redshift of \1:
[Si/H]=\gpm{-0.9}{0.6}{0.4}, [Fe/H]=$-1.3\pm0.3$, and
[Al/H]=\gpm{-1.0}{0.5}{0.3}. Assuming that the identification of
[\ion{O}{ii}]~\l 3727 in the \2\ spectrum is correct, we derive a
star-formation rate of $SFR_{\rm [OII]}=1.4$~M\subsun\ yr$^{-1}$,
which is a typical value for GRB host galaxies.

\begin{acknowledgements}
  PMV thanks Thomas Szeifert for an enlightening discussion on the
  flux-calibration of FORS spectra. The rapid follow-up VLT
  spectroscopic observations discussed in this paper would not have
  been possible without the GRB Coordinates Network (GCN), set up by
  Scott Barthelmy at NASA's GSFC. We are grateful to Jochen Greiner
  and the University of Texas GRBlog team led by Robert Quimby for
  their extremely useful GRB webpages. We acknowledge benefits from
  collaboration within the Research Training Network ``Gamma-Ray
  Bursts: An Enigma and a Tool'', funded by the EU under contract
  HPRN-CT-2002-00294. PMV was partly supported by the NWO Spinoza
  grant 08-0 to E.P.J. van den Heuvel. LK has been supported by a
  fellowship of the Royal Academy of Arts \& Sciences in the
  Netherlands.  JG acknowledges the support of a Ram\'on y Cajal
  Fellowship from the Spanish Ministry of Education and Science.  The
  Dark Cosmology Centre is supported by the DNRF.
\end{acknowledgements}

\bibliographystyle{aa}
\bibliography{references}

\end{document}